\documentclass[twocolumn,prc,aps,floatfix]{revtex4}
\usepackage[dvips]{graphicx}
\begin{document}
\title{Exchange interaction and correlations radically change behaviour of a quantum particle in a classically forbidden region}
\author{V.V. Flambaum}
\affiliation{
 School of Physics, The University of New South Wales, Sydney NSW
2052, Australia
}
\date{\today}
\begin{abstract}
Exchange interaction strongly influences the long-range behaviour of
localised electron orbitals and quantum tunneling amplitudes. It produces a power-law decay instead of the
usual exponential decrease at large distances. For inner orbitals inside
 molecules decay is $r^{-2}$, for macroscopic systems
 $\cos{(k_f r)} r^{-\nu}$, where $k_f$ is the Fermi momentum and 
   $\nu=3$ for 1D, $\nu=$3.5 for 2D and $\nu=$4 for 3D crystal.
Correlation corrections do not change these conclusions.
 Slow decay increases the exchange interaction
 between localised
 spins and the under-barrier tunneling amplitude.  The
under-barrier transmission
 coefficients in solids (e.g. for point contacts) become temperature-dependent.
\end{abstract}
\maketitle

\section{Introduction}
  One of the first famous results of Quantum Mechanics was that a particle
 may tunnel through a potential barrier. The tunneling amplitude
is exponentially small in the classical limit.
As we will see below this result may be incorrect if we take into account
the exchange interaction. The exchange interaction is described by
 the non-local (integration) operator, and the well-known theorems proven
 for the Schroedinger equation with a local potential $U({\bf r})$ are violated
if we add the exchange term (or any other non-local operator).
 A similar effect is produced by the
 correlation corrections. In this letter we consider the influence of the
 exchange interaction and correlations on an electron orbital in an atom, molecule or solid. 
The tunneling amplitude is still small in the classical limit, however
the decay of the orbitals in the classically forbidden area is much slower
 ($r^{-\nu}$) and depends on the dimensionality of the system.

The Hartree-Fock equation for an electron orbital  $\Psi({\bf r})$ in an atom, molecule or solid has the following form:
\begin{equation}\label{HF}
-\frac{\hbar ^2}{2m}\frac{d^2}{d{\bf r}^2} \Psi({\bf r}) +(U({\bf r})-E) \Psi({\bf r})= K({\bf r})
\end{equation}
\begin{equation}\label{HFK}
 K({\bf r})=\sum_q \Psi_q({\bf r}) \int \Psi_q({\bf r'})^\dagger \frac{e^2}{|{\bf r-r'}|} \Psi({\bf r'}) d {\bf r'}
\end{equation}
Here the summation runs over all electron orbitals  $\Psi_q({\bf r})$ with the same spin projection as  $\Psi({\bf r})$.
Now consider, for example, an inner electron atomic orbital 1$s$. The solution of the Schroedinger equation in 
potential $U({\bf r})$ has  a very small range $a_B/Z$ where $Z$ is the nuclear charge. Outside this range the orbital decreases exponentially
as $\exp{(-r Z/a_b)}$. In the Hartree-Fock equation (\ref{HF}) such rapid decay is impossible if an atom has more than two electrons.  Indeed, if 
  $ \Psi({\bf r}) \sim \exp{(-r Z/a_b)}$ the left-hand-side of  eq. (\ref{HF}) would be exponentially small while
the right-hand-side is still large since   $K({\bf r})$ in  eq. (\ref{HFK}) contains higher orbitals  $\Psi_q({\bf r})$
which have larger range. The behaviour of the inner Hartree-Fock orbitals inside  atoms have been studied analytically
(in the semiclassical approximation) and numerically in Ref. \cite{DFS}
 (see also section IIA).  The dependence on the radius $r$ can be found from
the multipole expansion of  $|1/({\bf r-r'})|$ in   $K({\bf r})$; the slowest decay normally comes from the dipole
term  $(\sim r'/r^2)$ and/or  last occupied orbital $\Psi_q({\bf r})$, $K \sim \Psi_q({\bf r})/r^2$. The extra nodes
 appear since the orbitals
  $\Psi_q({\bf r})$ oscillate.
For example, the 1$s$ orbital in Cs atom has 3 nodes  \cite{DFS} (without
 the exchange term a ground state has no nodes).
  The existence of extra nodes in solutions of Hartree-Fock equations
was also mentioned in Ref. \cite {Froese}.  Outside the atom all orbitals decay with exponential factor for an external
 electron \cite{Handy}.

 Inside solids there are electrons in the conduction band which occupy the whole crystal.
It has been pointed out in Ref. \cite{DFS} that the exchange interaction
 between localised bound electron and the conduction band electrons leads
 to a power tail of the bound electron orbital. 
 The effect of the exchange
 interaction $K({\bf r})$ has been estimated in the free band electron
 approximation
  $\Psi_q({\bf r}) = \exp{(i {\bf q \cdot r})}$.
An orbital of a bound electron decreases at large distances as  \cite{DFS}
\begin{equation}\label{tail}
  \Psi({\bf r}) \sim \cos{(k_f r)}/r^4
\end{equation}
 where $k_f$ is the Fermi momentum. Note that this solution does not
 contradict to the  Bloch's theorem since we consider localised bound electron
 (e.g. on an impurity atom) which does not belong to any electron band
in the periodic potential. It is curious that $1s$ orbital of an atom placed
 in a crystal has  infinite number of oscillations.

The derivation of this expression assumes the presence of a partly filled
 conduction electron band. However, in atoms and molecules of any length
the exchange enhancement of the inner orbital tail may be mediated by
 a complete electron shell.
 The question is: can the exchange enhancement in solids be mediated by
a nonconducting electron band? A special interest in this problem may be
 motivated by spintronics and solid state quantum computers based
 on spin qubits.  The long-range tail of the wave function  could,
 in principle,  lead to an enhancement of the exchange spin-spin
 interaction between the distant localised spins, and enhancement of the under-barrier tunneling amplitude. 

  A special feature of the ``long-tail'' mechanism is that
the state of the band electrons does not change, i.e.
  there is no need to have polarization of the conduction band by the localised
spin. The mediating band electrons produce the mean exchange field
 $K({\bf r})$ in  eq. (\ref{HFK}) only.
Therefore,  this ``long-tail'' effect is different from other 
 effects like the RKKI interaction \cite{RKKI} and 
 the double exchange  spin-spin  interaction suggested by Zener \cite{Zener}
 (see also Refs. \cite{Anderson,deGen} and description of Anderson and Kondo
 problems , e.g., in the book \cite{Phillips}).

 To investigate this problem in the present paper we perform calculation
 of the tail using the
Bloch waves and  tight-binding band electron wave functions.  

\section{Atom}
\subsection{Exchange}
Let us first explain how the long tail appears in atoms \cite{DFS}.
The radial equation for a Hartree-Fock electronic orbital
 $\xi_i(r)=r \phi_i(r)$ is  
\begin{equation}\label{rad}
[-\frac{\hbar^2}{2m}\frac{d^2}{dr^2}+(U_{eff}-E_i)]\xi_i(r)=K_i(r)
\end{equation}
\begin{equation}\label{Ueff}
U_{eff}=U+\frac{\hbar^2l(l+1)}{2mr^2} \ .
\end{equation}
The radial exchange term can be obtained using the multipole expansion of $1/|{\bf r-r'}|$. Outside the radius of an
inner orbital $\xi_i$ (e.g. in the area  $r> a_B/Z$ for $1s$) 
\begin{equation}\label{Ki}
K_i(r)=\sum_{k>0,n} C_{nk} b_{nk}\frac{\xi_n(r)}{r^{k+1}} \ .
\end{equation}
Here $C_{nk}$ are the standard angular momentum dependent coefficients and $b_{nk}=\int r^k \xi_n(r) \xi_i(r) dr$.
For the multipolarity $k=0$ the integral $b_{nk}=0$ due to the orthogonality of radial wave functions
with the same angular momentum. 

  Now we can discuss the large distance behaviour of the orbital $\xi_i(r)$. We will use 1$s$ orbital in  Xe atom ($Z=54$)
 as an example.  The last occupied shells are ...$5s^25p^6$. The orbital $5s$
  does not contribute
to $K_i(r)$ since in this case the multipolarity of the exchange integral is $k=0$ and the orthogonality condition
makes $b_{nk}=0$. The exchange integral $1s5p$ has $k=1$, therefore, at $r \sim a_B$ and outside the atom
$K_{1s}(r)\approx C_{5p,1} b_{5p,1} \xi_{5p}(r)/r^2$.

 The solution of Eq. (\ref{rad}) may be presented as \cite{comment}
\begin{equation}\label{xi}
\xi_i(r)=\xi^{free}_i(r)+\xi^{ind}_i(r)
\end{equation}
\begin{equation}\label{xiind}
\xi^{ind}_i(r)=[-\frac{\hbar^2}{2m}\frac{d^2}{dr^2}+(U_{eff}-E_i)]^{-1}K_i(r)
\end{equation}
Outside the radius of the inner orbital ($r>a_B/Z$ for $1s$) the energy $E_i$ is much larger than other
terms in the denominator of Eq. (\ref{xiind}) which are of the order of $E_n$ (since the opertor in the denominator acts
on $\xi_n$). In our example the energy
of $1s$ is $|E_i|=Z^2 \times 13.6$ eV=$4\cdot10^4$ eV while the $5p$ energy is $|E_n| \sim 10$ eV.    
Therefore, we can approximately write 
\begin{equation}\label{xiind1}
\xi^{ind}_i(r)=\frac{K_i(r)}{U_{eff}-E_i} +\frac{\hbar^2}{2m(U_{eff}-E_i)}\frac{d^2}{dr^2}\frac{K_i(r)}{U_{eff}-E_i}+...
\end{equation}
The free solution in this area may be described by the semiclassical (WKB) approximation, 
 \mbox{$\xi^{free}_i(r)\sim |p|^{-1/2}\exp{(-\int |p|dr/\hbar)}$};
 it has the usual range $a_B/Z=0.02 a_B$ for $1s$. Comparison with the numerical
solution of the Hartree-Fock equation for $1s$ orbital has shown that within $\sim$1\%  accuracy
  it is enough to keep the first two terms in the expansion Eq. (\ref{xiind1})  beyond the classical turning point,
  and only one term at $r>10a_B/Z$.
Similar results have been obtained for the Dirac-Hartree-Fock orbitals which include the spin-orbit interaction
and other single-particle relativistic corrections \cite{DFS}. Thus we see that at large distances
\mbox{$\xi_{1s}(r)\approx \rm{const}\, \xi_{5p}(r)/r^2$}.  

\subsection{Correlations}
The effect of the correlations may be described by the non-local ``correlation potential''
$\Sigma(r,r',E)$ (integration operator) which  modifies  electron orbitals (see e.g. \cite{DFSS1987}).
 The correlation potential is defined such 
that its average value coincides with the correlation correction to 
the energy,
\begin{eqnarray}
\delta E_i&=&\langle i |\hat{\Sigma}|i \rangle\\
C({\bf r_2})\equiv \hat{\Sigma}\Psi _i&=&\int\hat{\Sigma}({\bf r}_1,{\bf r}_2,E_i)
\Psi_i({\bf r}_1)d^{3}r_{1} \ .
\end{eqnarray}
By solving the Hartree-Fock equation for the  electron orbital including 
the correlation potential $\hat{\Sigma}$, we obtain ``Brueckner'' 
orbitals and energies:
\begin{equation}\label{BHF}
-\frac{\hbar ^2}{2m}\frac{d^2}{d{\bf r}^2} \Psi({\bf r}) +(U({\bf r})-E) \Psi({\bf r})= K({\bf r})+ C({\bf r})
\end{equation}
It is easy to write the correlation potential explicitly. In the  
second-order perturbation theory in the residual interaction there are four term.
The direct term $\hat{\Sigma}^{d}({\bf r}_1,{\bf r}_2, E_i)$ is
 given by
\begin{equation}\label{sigmad} 
e^4\sum_{n,\beta,\gamma}\int
d{\bf r_3}d{\bf r_4}\frac{\psi _{n}^{\dag}({\bf r}_{4})\psi _{\beta}({\bf r}_{4})\psi _{\gamma}({\bf r}_{2})
\psi _{\beta}^{\dag}({\bf r}_{3})\psi ^{\dag}_{\gamma}({\bf r}_{1})
 \psi _{n}({\bf r}_{3})}
{r_{24}r_{13}
(E_i +\epsilon_{n}-\epsilon_{\gamma}-\epsilon_{\beta})} .
\end{equation}
Note that $\hat{\Sigma}^{d}$ is a single-electron and energy-dependent 
operator. At large distance this term becomes the well-known local polarization
potential $\sim 1/r^4$
 (see e.g. \cite{DFSS1987}),
 so it is not interesting for us.
An interesting contribution comes from the exchange correlation potential
  $\hat{\Sigma}^{exch}$
\begin{equation}\label{sigmaex}
e^4\sum_{n,\beta,\gamma}\int 
d{\bf r_3}d{\bf r_4}\frac{\psi _{n}^{\dag}({\bf r}_{4})\psi _{\beta}({\bf r}_{4})\psi _{\gamma}({\bf r}_{2})
\psi _{\beta}^{\dag}({\bf r}_{1})\psi ^{\dag}_{\gamma}({\bf r}_{3})
 \psi _{n}({\bf r}_{3})}
{r_{24}r_{13}
(E_i +\epsilon_{n}-\epsilon_{\gamma}-\epsilon_{\beta})}.
\end{equation}
 In this case we have the situation similar to the exchange interaction.
Consider, for example,  the correlation correction to the Xe $1s$ orbital ($i=1s$) and $n=5p$. The energy of $1s$
is large and to make an estimate we can neglect $\epsilon_{5p}-\epsilon_{\gamma}-\epsilon_{\beta}$ in the denominator
of the Eq.  (\ref{sigmad}). After the summation over $\beta$ and $\gamma$ we
 obtain the exchange correlation term
 $C^{exch}({\bf r})\sim 2 e^2/(rE_{1s})K({\bf r})$ where  $K({\bf r})$ is the
 usual exchange term.
Therefore, at large $r$ the correlation term is suppressed in comparison
 with the exchange term by the small factor $2 e^2/(rE_{1s})$.  
For $1s$ orbital at  $r=a_B$  the suppression factor  is $4/Z^2$.
The correlations are more important for higher orbitals where
 the suppression factor  is  $2 e^2/(rE_{i})$ and  $|E_{i}| \ll |E_{1s}|$.

We may conclude that within the perturbation theory treatment
the  correlations do not produce qualitative changes in the properties of
the long-range tail. Their effect is similar to that of the exchange, however,
the decay is faster (extra $1/r$).

\section{1D, 2D and 3D systems}
  If we consider a molecule instead of atom, inner electron orbital will behave
the same way,  \mbox{$\xi_{inner}(r)\approx \rm{const}\, \xi_{valence}(r)/r^2$}.
In macroscopic systems there is a large number of electrons occupying the
valence band and the contribution of different valence electrons interfere
in the exchange term in Eq.(\ref{HFK}). This interference changes the
 long range behaviour.

The equation for a bound electron wave function $\Psi_b({\bf r})$ in a crystal
contains the exchange term from Eq.(\ref{HFK}) describing the exchange interaction of the bound electron
with $2F$  mobile electrons:
\begin{equation}\label{K1}
 K({\bf r})=\int g({\bf r}-{\bf r}')
 [\frac{e^2}{|{\bf r}-{\bf r}'|}-\frac{e^2}{r}] \Psi_b({\bf r}') d {\bf r}',
\end{equation}
\begin{equation}\label{g}
g({\bf r}-{\bf r}')  \equiv \sum_n \Psi_n({\bf r})  \Psi_n({\bf r}')^\dagger .
\end{equation}
Summation goes over F mobile electron  states $\Psi_n({\bf r})$ with the same spin projection.
 To account for the orthogonality condition $\int  \Psi_n({\bf r}')^\dagger\Psi_b({\bf r}') d {\bf r}'=0$
in Eq. (\ref{K1}) we  excluded the zero multipolarity term from the Coulomb integrals, replacing 
$\frac{e^2}{|{\bf r}-{\bf r}'|}$
 by $\frac{e^2}{|{\bf r}-{\bf r}'|}- \frac{e^2}{ r}$.
In the  ``exact'' expression (\ref{K1})  the subtracted term
 $ \frac{e^2}{|{\bf r}|}$ disappears after
 the integration over ${\bf r'}$ since $\int  \Psi_n({\bf r}')^\dagger\Psi_b({\bf r}') d {\bf r}'=0$.

Let us start discussion of crystals
from the simplest problem - a 1D chain of $N$ atoms separated by distance  $a$.
The wave function of a mobile electron can presented as
\begin{equation}\label{mobb}
\Psi_n({\bf r})=L^{-1/2}  e^{ik_n x} v_k({\bf r}),
\end{equation}
where $v_k({\bf r})$ is a periodic function in $x$-direction and $L=Na$ is the length of the chain. 
To perform the summation in Eq. (\ref{g}) analytically we neglect dependence on $k$ in $v_k({\bf r})$.
Taking the standard set of the wave vectors $k_n=2 \pi n/L$, $n=0, \pm 1,...,\pm q$, where $F=2q+1$, we obtain
\begin{equation}\label{gk}
g({\bf r}-{\bf r}')=  v({\bf r})v({\bf r}') \frac{\sin{[k_f (x-x')] }}{x-x'} .
\end{equation}
where $k_F=f \pi/a $ and $f=F/N$ is the band filling factor. Now we can find the exchange term Eq (\ref{K1}).
The leading term in the multipole expansion ($r'<<r$) of \mbox{$\frac{e^2}{|{\bf r}-{\bf r}'|}- \frac{e^2}{ r} \approx\frac{e^2 ({\bf r \cdot r'})}{r^3} $} leads to the
dipole approximation for  $K({\bf r})$ at large distance:
\begin{eqnarray}\label{K2}
\nonumber
 K({\bf r})=\frac{e^2v({\bf r})}{\pi r^3}[\sin{(k_fx)}\int x'\cos{(k_fx')} v({\bf r}')  \Psi_b({\bf r}') d {\bf r}'\\
-\cos{(k_fx)}\int x'\sin{(k_fx')} v({\bf r}')  \Psi_b({\bf r}') d {\bf r}']
\end{eqnarray}

 It is easy to extend the problem
 to 2$D$ and 3$D$ cases. In 2D case we obtain
\begin{equation}\label{gk2}
g({\bf r}-{\bf r}')=  v({\bf r})v({\bf r}') \frac{J_1(k_f R)}{2 \pi R}
\sim \frac{\sin{(k_f R-\pi/4)}} {R^{3/2}} \,
\end{equation}
 where ${\bf R}={\bf r}-{\bf r}'$ and $J_1$ is the Bessel function.
In 3D case
\begin{equation}\label{gk3}
g({\bf r}-{\bf r}')= 
 \frac{ v({\bf r})v({\bf r}') }{2 \pi^2 R^2 }[-\cos{(k_f R)}
+\frac{\sin{(k_f R)}} {k_f R}] .
\end{equation}
Substituting these results into  Eq (\ref{K1}) we obtain in
 the dipole approximation that the exchange
 interaction term decays as
\begin{equation}\label{K3}
 K(r) \sim \cos{(k_f r)} r^{-\nu} \,
\end{equation}
 where $k_f$ is the Fermi momentum and 
   $\nu=3$ for 1D, $\nu=$3.5 for 2D and $\nu=$4 for 3D crystal,
 i.e. $\nu=(5+d)/2$ where $d=1,2,3$ is the dimension.

 The long-range tail in  Eq (\ref{K3}) is due to the exchange interaction
between bound electrons and  conducting electrons which travel
freely inside the crystal and may be found at any distance from
the bound electron. As we have seen in section II, the perturbation theory treatment
of the correlations does not change our conclusions. This is the normal metal case
where the correlations are realatively weak.
In this case the long-range tail of a bound electron orbital
is the real physical phenomenon which should be taken into account,
for example, in calculating tunneling amplitudes or exchange
interaction between distant localised spins.

 Note that the expressions
 (\ref{K2},\ref{K3}) do not vanish if the electron band is complete.
Instead they  have fast oscillations if the electron Fermi momentum  $k_f$ is
 large. This conclusion looks surprising since a complete band does not
 contribute to the  conductivity.
 One may compare this crystal complete band case with a molecule where 
 valence electrons present on all atoms even in the absence of the conductivity.
Therefore,  one may have, in principle, an enhanced tunneling
 amplitude or enhanced exchange interaction
between distant spins (power suppression $r^{-\nu}$ instead of exponential
 suppression) even in non-conducting materials.
However, if there are  strong electron-electron correlations
(due to the strong repulsion between valence electrons located at the
 same site), they
 transform the  Bloch-Hartree-Fock  (conductor) state into the Mott insulator
 state where there are no free electrons and no long tail.
 
 The long-tail effect  does not appear in any approach where
the exchange interaction is replaced by an effective  potential or by a
density-dependent potential.
Approximate calculations may also lead to other incorrect conclusions.
For example,  the long-range tail for a complete band case does not appear
  in the tight-binding approximation for the electron wave functions.
 In the tight-binding approximation a wave function of mobile electron is
\begin{equation}\label{mob1}
\Psi_n({\bf r})=N^{-1/2} \sum_l e^{ik_n l a} \Psi_1({\bf r}-l{\bf a}),
\end{equation} 
where $\Psi_1({\bf r}-l{\bf a})$ is the one-site wave function.
 The substitution of $\Psi_n$ from Eq. (\ref{mob1}) into  Eq. (\ref{g})
and summation over $n$ gives the following results:
\begin{equation}\label{sum}
g({\bf r}-{\bf r}')=\sum_{l,m}B(F,l-m)\Psi_1({\bf r}-l{\bf a})
\Psi_1({\bf r'}-m{\bf a}) ^\dagger
\end{equation}
\begin{equation}\label{B}
B(F,l)=\frac{\exp{(i2\pi lF/N)}-1}{N (\exp{(i2\pi l/N)}-1)}\approx \frac{\exp{(i\pi f l)}}{ \pi l}\sin{(\pi fl)},
\end{equation}
where $l>0$, $f=F/N$ is the band filling factor and the last expression is obtained for $l \ll N$. For 
$l=0$ we have $B(F,0)=f$.
Substitution of $g({\bf r}-{\bf r}')$ from Eq. (\ref{sum}) into 
 Eq. (\ref{K1}) shows that  if the band is partly
filled, the tight-binding approximation leads to the same conclusion
$ K(r) \sim \cos{(k_f r)} r^{-\nu}$.
However, for the  completely filled band $f=1$ and $\sin{(\pi fl)}=0$.
This means that the long-range exchange term vanishes in the absence
 of mobile carriers, electrons or holes. The explanation is simple:
 in the tight-binding approximation the
 complete band wave function made of the running waves Eq. (\ref{mob1})
 is equal to the
 antisymmetrised  product of the localised electron wave functions
 $\Psi_1({\bf r}-l{\bf a})$. The exchange interaction with the localised
electrons does not produce the long-range tail. To compare with the Bloch wave
expression one may say that the tight-binding result for the complete
band  corresponds to $K(r) \sim \sin{(k_f r)}=0$ for $r=la$. However,
the  oscillations  of $K(r)$ do not lead to vanishing of its effect on the wave
 functions  - compare with the solution for atomic orbitals in the previous
section.

  At finite temperature  conducting electrons and holes appear. This activates
the long-tail mechanism even in the tight-binding approximation
 and makes the under-barrier transmission coefficient
 temperature dependent. Here it may be appropriate to recall
 that a temperature dependence of the transmission
coefficient has been observed near the  ``0.7 $(2e^2/h)$ structure''
 in the point contact conductance measurements \cite{Thomas96,Thomas}.

This work is  supported by the Australian Research
Council.


\begin{thebibliography}{99}
\bibitem{comment} The dependence of $K_i$ on $\xi^{ind}$ is not important
 since the integral
 $b_{nk}=\int r^k \xi_n(r) \xi_i(r) dr$ is convergent at small distances where
  $\xi^{ind}\ll\xi^{free}$.
\end{thebibliography}
\end{document}